\documentclass[12pt]{article}
\usepackage{epsfig,amsmath,amsthm,amssymb,graphics,amsfonts,psfrag}
\usepackage[letterpaper,margin=1in]{geometry}

\theoremstyle{plain}

\theoremstyle{definition}

\def\R{\mathbb{R}}

\newcommand{\be}{\begin{equation}}
\newcommand{\ee}{\end{equation}}
\newcommand{\bea}{\begin{eqnarray}}
\newcommand{\eea}{\end{eqnarray}}
\newcommand{\beann}{\begin{eqnarray*}}
\newcommand{\eeann}{\end{eqnarray*}}
\newcommand{\benn}{\begin{equation*}}
\newcommand{\eenn}{\end{equation*}}

\def\ra{\rightarrow}

\begin{document}
 
\title{On spatial and temporal multilevel dynamics and scaling effects in epileptic seizures}
\author{Christian Meisel\thanks{Max Planck Institute for the Physics of Complex Systems, N\"{o}thnitzer Stra\ss e 38, 01187 Dresden, Germany}~\thanks{Department of Neurology, University Clinic Carl Gustav Carus, Fetscherstra\ss e 74, 01307 Dresden, Germany}~ \& Christian Kuehn\thanks{Max Planck Institute for the Physics of Complex Systems, N\"{o}thnitzer Stra\ss e 38, 01187 Dresden, Germany}~\thanks{equal contribution}}

\maketitle

\begin{abstract}
Epileptic seizures are one of the most well-known dysfunctions of the nervous system. During a seizure, a highly synchronized behavior of neural activity is observed that can cause symptoms ranging from mild sensual malfunctions to the complete loss of body control. In this paper, we aim to contribute towards a better understanding of the dynamical systems phenomena that cause seizures. Based on data analysis and modelling, seizure dynamics can be identified to possess multiple spatial scales and on each spatial scale also multiple time scales. At each scale, we reach several novel insights. On the smallest spatial scale we consider single model neurons and investigate early-warning signs of spiking. This introduces the theory of critical transitions to excitable systems. For clusters of neurons (or neuronal regions) we use patient data and find oscillatory behavior and new scaling laws near the seizure onset. These scalings lead to substantiate the conjecture obtained from mean-field models that a Hopf bifurcation could be involved near seizure onset. On the largest spatial scale we introduce a measure based on phase-locking intervals and wavelets into seizure modelling. It is used to resolve synchronization between different regions in the brain and identifies time-shifted scaling laws at different wavelet scales. We also compare our wavelet-based multiscale approach with maximum linear cross-correlation and mean-phase coherence measures. 
\end{abstract}


\newpage
\section{Introduction}  
\label{sec:intro}

Trying to predict epileptic seizures using time series analysis has been an important research topic for decades. In particular, the now wide-spread use of EEG (electroencephalography) techniques to acquire data has been a major driving force of the subject. The review article \cite{MormannAndzejakElgerLehnertz} and the recent book \cite{SchelterTimmerSchulze-Bonhage} provide perspectives what has been achieved in seizure prediction. The main goal was to identify and characterize a pre-ictal phase occurring before the onset and to design measures that approximately predict the critical starting time of the seizure \cite{LittEchauz}. Since research has focused in this direction there are still gaps \cite{RobinsonRennieRowe} in our understanding of seizures from a dynamical systems perspective \cite{Wendling,LopesdaSilvaetal}. In this paper, we are going to address this issue and focus on dynamical mechanisms \cite{ShustermanTroy} instead of aiming at a predictive technique for seizures.

The main themes of our results are the deep links to mathematical multiscale techniques \cite{Rinzel,PercivalWalden} and the observation of scaling laws at different spatio-temporal levels. From models based on biophysical principles of brain dynamics it is expected that multiple spatial \cite{BreakspearStam} and multiple time scales \cite{HoneyKoetterBreakspearSporns} play an important role for epileptic seizures \cite{Richardson}. Based on a combination of analyzing epileptic seizure patient data and neuron modelling we split the problem into three spatial scales and show that at each individual spatial level the problem exhibits multiple time scale behaviour. We point out that our approach to verify the existence of multiscale phenomena is primarily data-driven and complements modelling approachs (see e.g. \cite{Decoetal}).\\ 

On the smallest spatial scale, we employ model-based analysis of single neurons \cite{Izhikevich1,KeenerSneyd1} using a multiple time scale stochastic FitzHugh-Nagumo model \cite{FitzHugh,Nagumo,Lindneretal} with a focus on early-warning signs \cite{Schefferetal}, scaling laws and control failure of spiking. In particular, we investigate three different cases of spiking and provide the first results of critical transition theory \cite{KuehnCT1} for neurons in an excitable state. Critical transition theory for systems without equlibria near bifurcations has recently been applied successfully in climate modeling \cite{Lentonetal,Alleyetal} and in ecological systems \cite{Clarketal,Carpenteretal} but apparently has not been applied to neuroscience problems yet. We analyze three different regimes for the relationship between noise and time scale separation and show that the variance can be a precursor of spiking in some parameter regimes while it fails in the low noise case. In this context, we point out that the distributions of interspike intervals \cite{Lindner} has been studied extensively in single neuron models but that our work only studies the time series locally near a bifurcation and does not require multiple events.  

The second spatial scale which we consider are clusters/regions of neurons \cite{Osorioetal,Schelteretal}. Here we use electrocorticogram (ECoG) data; see Appendix \ref{ap:data}. We examine the onset of the epileptic seizure using the variance as a simple univariate measure. We observe that during a certain period before the seizure the variance shows oscillations. Furthermore, very close to the transition to a seizure the inverse of the variance displays a linear scaling law. Based on critical transition theory, these observations are generically characteristics for Hopf bifurcation \cite{KuehnCT2}. It is very important to note that many seizure models \cite{Rodriguesetal,Rodriguesetal1,Martenetal,SuffczynskiKalitzinLopesdaSilva,Breakspearetal} suggest a Hopf bifurcation as a main mechanism as the transition point. Therefore, our results not only provide a first application of local scaling laws near bifurcations to data but also validate the proposed bifurcation mechanism arising from biophysical principals. Similar to the individual neurons scale, we point out that distributions of interseizure intervals have been studied \cite{SuffczynskiKalitzinLopesdaSilva} but that we do not require multiple events.
 
On the largest spatial scale we analyze the synchronization and correlation between different brain regions \cite{Lehnertzetal}. Several bivariate measures have been proposed \cite{MormannAndzejakElgerLehnertz} to study epileptic seizures but the underlying complex network structure makes the problem difficult \cite{KuhnertElgerLehnertz}. Our approach utilizes a recent technique calculating phase-locking intervals (PLIs) \cite{KitzbichlerSmithChristensenBullmore} based on wavelet transforms \cite{PercivalWalden}. Wavelet-based methods have been applied previously in the context of epileptic seizures \cite{Bosnyakovaetal} but our approach is the first to investigate PLIs and associated phase-locking. We show that our wavelet-based method \cite{KitzbichlerSmithChristensenBullmore,PercivalWalden} measures increasing phase-locking and resolves a multiple time scale structure near the seizure onset. Furthermore, we observe a linear scaling law of average phase-locking and that phase-locking at different scales often starts at different times where the lower scales tends to increase earlier compared to higher scales. These results apply near the seizure onset and could potentially relate to recently observed rapid discharges \cite{Wendlingetal,Molaee-ArdekaniBenquetBartolomeiWendling}. We also compare our results to other bivariate measures such as maximum linear cross-correlation \cite{RosenblumPikovskyKurths,Feldwisch-Drentrupetal} and mean phase coherence \cite{Chavezetal}.\\

In summary, our study introduces two recently developed methods (critical transitions, wavelets/PLIs) into the analysis of epileptic seizures. Using critical transitions theory we give the first analysis of early-warning signs for excitable neurons, identify a potential Hopf bifurcation as the seizure onset mechanism from data and find a new scaling law of single-event time series data at the cluster level. For the wavelet-based phase-locking technique, we provide a comparative study to other bivariate measures and discover a scaling law occurring at time-shifted onset times. On each of the three spatial levels we were also able to identify a multiple time scales structure, based on a data-driven time series approach.       

\section{Single Neurons}

We start on the level of single neurons. Clearly it is very problematic to get data in this case before epileptic seizures so that we resort to model neurons. The main question will be whether we can predict a spike in the voltage time trace of the model neuron before it occurs. The FitzHugh-Nagumo (FHN) model \cite{FitzHugh,Nagumo,GGR} is a simplification of the Hodgkin-Huxley equations \cite{HodgkinHuxley} which model the action potential in a neuron. We point out that the methods we are going to present here are going to apply to a much wider class of excitable neuronal models than the FHN equation such as the original Hodgkin-Huxley model \cite{RubinWechselberger1} or the Morris-Lecar system \cite{GuckenheimerKuehn2} since these models have similar bifurcation structure and multiple time scale properties \cite{Izhikevich1}.\\ 

\begin{figure}[htbp]
\centering
 \includegraphics[width=1\textwidth]{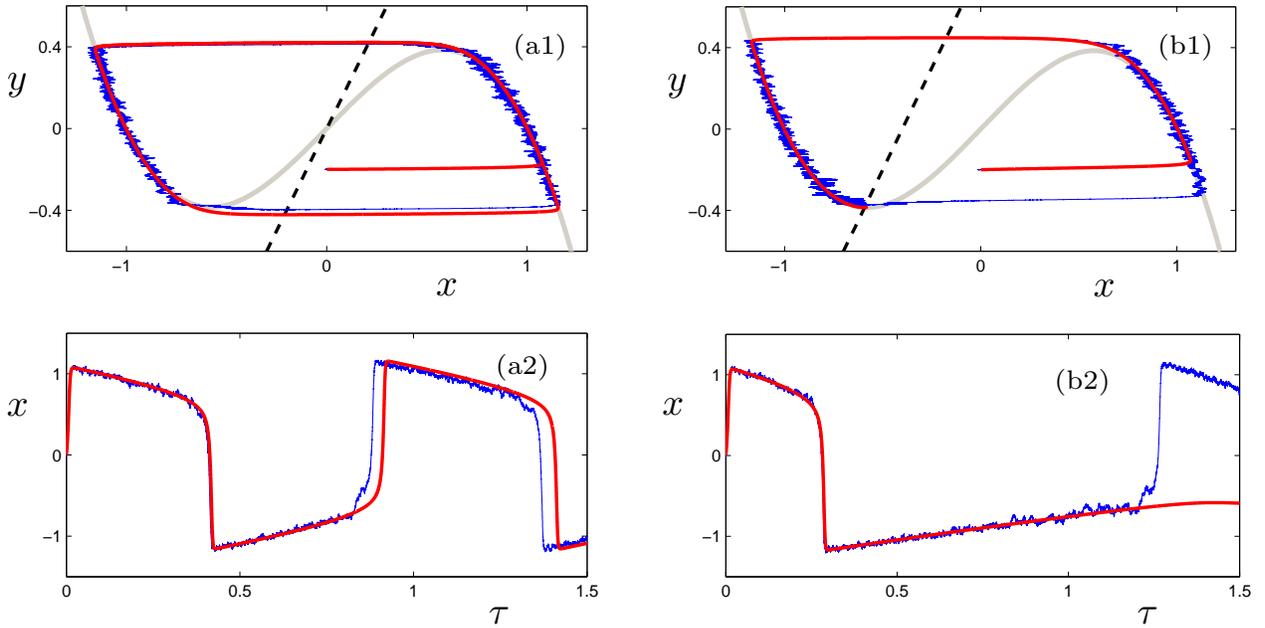}
\caption{\label{fig:fig1}Simulation of \eqref{eq:FHN} with $\epsilon=0.005$ and $\gamma=2$ using an Euler-Maruyama numerical SDE solver \cite{Higham}; red curves are deterministic trajectories with $\sigma=0$ and blue curves are sample paths with $\sigma=0.0028$. Systems have always been started at $(x_0,y_0)=(0,-0.2)$. The critical manifold $C_0$ is shown in grey and the $y$-nullcline as a dashed black curve. (a) $b=0$, the equilibrium for the full system lies on $C_0^r$. (b) $b=0.8$, the equilibrium lies on $C_0^{a-}$ near the fold point $(x_-,y_-)$. The deterministic trajectory has only one spike while noise-induced escapes produce repeated spiking for the stochastic system.}
\end{figure}

There are several forms of the FHN-equation \cite{GuckenheimerKuehn1}. One possible version suggested by FitzHugh is the Van der Pol-type \cite{vanderPol} model
\be
\label{eq:FHN_det}
\begin{array}{rcrcl}
\epsilon\frac{dx}{d\tau}&=&\epsilon \dot{x} &=& x-x^3-y,\\
\frac{dy}{d\tau}&=&\dot{y}&=& \gamma x-y+b,\\
\end{array}
\ee
where $x$ represents voltage, $y$ is the recovery variable and $\gamma$, $b$, $\epsilon$ are parameters. We think of $b$ as an external signal or applied current \cite{LindnerSchimansky-Geier} and assume that the time scale separation $\epsilon$ satisfies $0\leq \epsilon\ll1$ so that $x$ is the fast variable and $y$ the slow variable. The dynamics of \eqref{eq:FHN_det} can be understood using a fast-slow decomposition \cite{Desrochesetal,MisRoz,Grasman}. Setting $\epsilon=0$ in \eqref{eq:FHN_det} gives a differential equation on the slow time scale $\tau$ defined on the algebraic constraint
\benn
C_0:=\{(x,y)\in\R^2:y=x-x^3=:c_0(x)\}.
\eenn  
We call $C_0$ the critical manifold; see Figure \ref{fig:fig1}. Differentiating $y=c_0(x)$ implicitly with respect to $\tau$ we find $\dot{y}=\dot{x}(1-3x^2)$ so that the differential equation on $C_0$ can be written as
\benn
\dot{x}=\frac{\gamma x-x+x^3+b}{1-3x^2}
\eenn
which we refer to as slow flow. Observe that the slow flow is not well-defined at the two points $(x_\pm,y_\pm)=(\pm x^{-1/3},c_0(\pm x^{-1/3}))$. Applying a time re-scaling to the fast time $t:=\tau/\epsilon$ to \eqref{eq:FHN_det} gives 
\be
\label{eq:FHN_det_fast}
\begin{array}{rcrcl}
\frac{dx}{dt}&=&x' &=& x-x^3-y,\\
\frac{dy}{dt}&=&y'&=& \epsilon(\gamma x-y+b).\\
\end{array}
\ee

Setting $\epsilon=0$ in \ref{eq:FHN_det_fast} gives the fast flow where $y'=0$ implies that $y$ is viewed as a parameter in this context. Observe that $C_0$ consists of equilibrium points for the fast flow and that the points $(x_\pm,y_\pm)$ are fold (or saddle-node) bifurcation points \cite{Strogatz} in this context. The critical manifold naturally splits into three parts
\benn
C_0^{a-}:=C_0\cap\{x<x_-\},\qquad C_0^{r}:=C_0\cap\{x_-<x<x_+\}, \qquad C_0^{a+}:=C_0\cap\{x>x_+\}
\eenn
where $C_0^{a\pm}$ are attracting equilibria and $C_0^r$ are repelling equilibria for the fast flow. We view $C_0^{a-}$ as the refractory state and $C^{a+}_0$ as the excited state for the neuron. For $\epsilon=0$ trajectories are concatenations of the fast and slow flows. We will consider two different situations for the parameters $(\gamma,b)$. In the first situation we chose the parameters so that \eqref{eq:FHN_det} has a single equilibrium point on $C^r_0\cap \Gamma$ where $\Gamma:=\{(x,y)\in\R^2:y=\gamma x+b\}$ is the $y$-nullcline of the FHN-equation; see Figure \ref{fig:fig1}(a1)-(a2). For $\epsilon=0$ suppose that $(x_0,y_0):=(x(0),y(0))\in C_0^{a-}$; then the slow flow moves the system to $(x_-,y_-)$, a jump via the fast subsystem to $C_0^{a+}$ occurs, the slow flow on $C^{a+}_0$ brings the system to $(x_+,y_+)$ and another jump returns it to $C_0^{a-}$. This is the classical relaxation oscillation \cite{Grasman,Guckenheimer6}. However, in neuroscience one often also considers the excitable regime \cite{Izhikevich1} where the global equilibrium $(x^*,y^*)$ for the system is stable and lies on $C_0^{a-}$ close to $(x_-,y_-)$; see Figure \ref{fig:fig1}(b1)-(b2). In this case, a trajectory of \eqref{eq:FHN_det} can generate, depending on $(x_0,y_0)$, at most one excursion/spike to the excitable state before returning to $(x^*,y^*)$. Repeated  spiking in the excitable regime can be obtained using the more general stochastic FHN-equation
\be
\label{eq:FHN}
\begin{array}{rcl}
\epsilon \dot{x}_\tau &=& x_\tau-x_\tau^3-y+\sigma \xi_\tau,\\
\dot{y}_\tau&=& \gamma x_\tau-y_\tau+b,\\
\end{array}
\ee
where $\xi_\tau$ is delta-correlated white noise $\langle\xi_{\tau_1}\xi_{\tau_2}\rangle=\delta(\tau_1-\tau_2)$ and $\sigma$ is a parameter representing the noise level. We can now ask whether individual neuron spiking activity already has precursors. This viewpoint should provide new insights how neurons are able to control synchronization and how control failure occurs. Recent results on predicting critical transitions \cite{Schefferetal} suggest that statistical precursors can be used to predict events similar to spiking in neurons from a time series without knowing their exact location. The detailed mathematical theory can be found in \cite{KuehnCT1,KuehnCT2}. 

Here we present the first application of this theory in the context of single neurons. We want to predict a spiking transition from a neighborhood of $C_0^{a-}$ to $C_0^{a+}$ and consider the variance as an early-warning sign
\benn
V_s:=\text{Var}(x_s)\qquad \text{restricted to $x_s$ near $C_0^{a-}$.}
\eenn 

\begin{figure}[htbp]
\centering
 \includegraphics[width=1\textwidth]{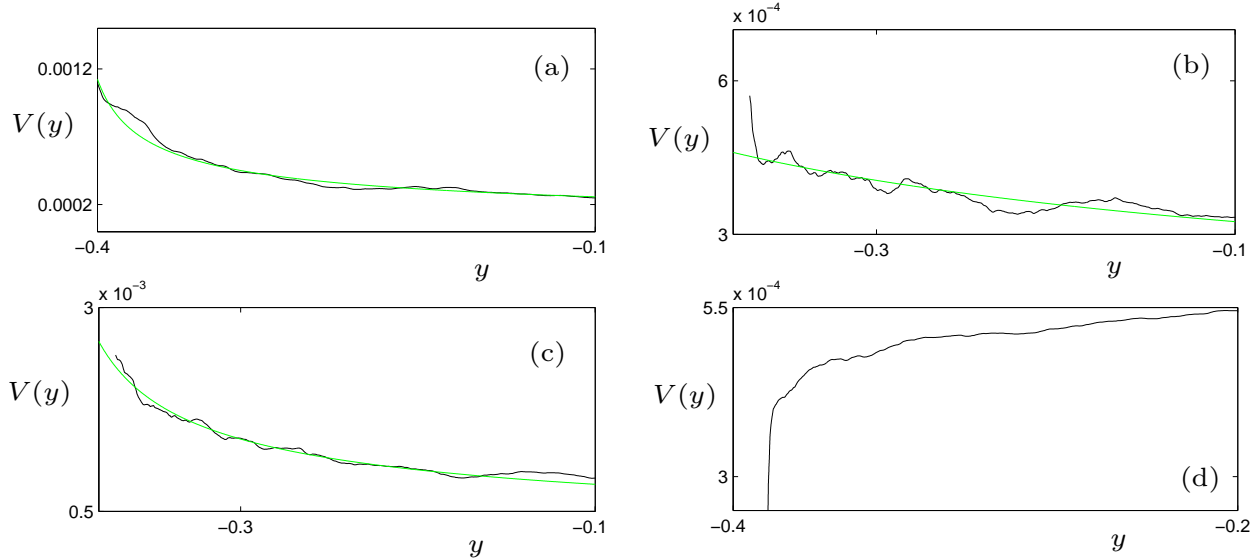}
\caption{\label{fig:fig2}Average of the variance $V(y)=\text{Var}(x(y))$ (black curves) over 100 sample paths starting for $s=0$ at $(x_0,y_0)=(-1,0)$ up to a final time $T$. The green curves are fits of $V$ using \eqref{eq:Var} with fitting parameters $A$ and $y_{\epsilon,-}$. Fixed parameter values are $(\epsilon,\gamma)=(2,0.005)$. (a) Relaxation oscillation regime with $(b,\sigma,T)=(0,0.02\sqrt\epsilon,0.28)$. (b) Excitable regime with $(b,\sigma,T)=(0.8,0.02\sqrt\epsilon,0.7)$; sample paths can exhibit oscillations around the stable focus equilibrium $(x^*,y^*)$ which are visible in the variance. (c) Excitable regime with $(b,\sigma,T)=(0,0.05\sqrt\epsilon,0.8)$ where larger noise regularizes the variance similar to (a). (d) Excitable regime $(b,\sigma,T)=(0,0.005\sqrt\epsilon,0.9)$ where smaller noise does not allow fast escapes from $(x^*,y^*)$ and yields decreasing variance.}
\end{figure}

Observe that we can view $V_s$ also as a function of $y$, and write $V=V(y)$, since the mapping between $y_s$ and $s$ is bijective when restricting to $C_0^{a-}$. In the relaxation oscillation regime (see Figure \ref{fig:fig1}(a1)-(a2)) and if $(\epsilon,\sigma)$ are sufficiently small it can be shown \cite{KuehnCT2,BerglundGentz4} that
\be
\label{eq:Var}
V(y)\sim \frac{A}{\sqrt{y-y_{\epsilon,-}}},\qquad \text{as $y\ra y_{\epsilon,-}$}
\ee
for some constant $A>0$ and where $|y_{\epsilon,-}-y_-|$ is small. Therefore an increase in fast voltage-variable variance can potentially be used to predict and to control spiking if no equilibrium exists near $(x_-,y_-)$. Here we extend the results of \cite{KuehnCT2} by investigating the excitable regime. Figure \ref{fig:fig2} shows an average of the variance $V$ computed over 100 sample paths using a sliding window technique \cite{KuehnCT1}. Figure \ref{fig:fig2}(a) shows the relaxation oscillation regime where we can confirm the theoretical prediction \eqref{eq:Var}.\\ 

The excitable regime is much more interesting since the equilibrium point $(x^*,y^*)$ can lead to a variety of distinct regimes depending on the noise level. In Figure \ref{fig:fig2}(b) the noise is at an intermediate level so that deterministic oscillations around the equilibrium are visible in the variance before an escape; hence the prediction \eqref{eq:Var} is not a good prediction of a spike but one should rely on the oscillatory mechanism before escapes. In Figure \ref{fig:fig2}(c) the noise is larger which provides a regularizing effect for the variance via noise-induced escapes. This relates to the well-known mechanism of coherence resonance \cite{Lindneretal}. In Figure \ref{fig:fig2}(d) the noise is very small so that sample paths need exponentially long times to escape and are metastable near $(x^*,y^*)$. This causes a decrease in variance and will make predictions very difficult. The different scaling regimes for noise level and time scale separation are discussed in more detail in \cite{DevilleVanden-EijndenMuratov,MuratovVanden-EijndenE,MuratovVanden-Eijnden,KuehnCT2}.\\ 

Based on our results we can conclude that control of spiking could depend crucially on noise level and statistical properties of the state of a neuron. In particular, in the excitable state already a small change in the noise level or system parameters can result in a substantial loss of control due to unpredictable spiking. This could cause undesirable synchronization and continuous spiking. Let us point out that this is just one possible explanation for a potential prediction/control failure during epileptic seizures but our results show that prediction at neuronal level can already be extremely complicated. Therefore we proceed to look at the next scale in our analysis and move from single neurons to clusters/regions of neurons.

\section{Local Data \& Clusters}
\label{sec:variance_local}

On the level of regions, we can start to analyze data obtained before epileptic seizures. The eight time series we use are described in detail in Appendix \ref{ap:data}. A natural extension of our previous strategy is to compute the variance for each time series using a sliding window technique and to understand the scaling laws associated with the variance on the cluster level.\\

\begin{figure}[htbp]
\centering
 \includegraphics[width=1\textwidth]{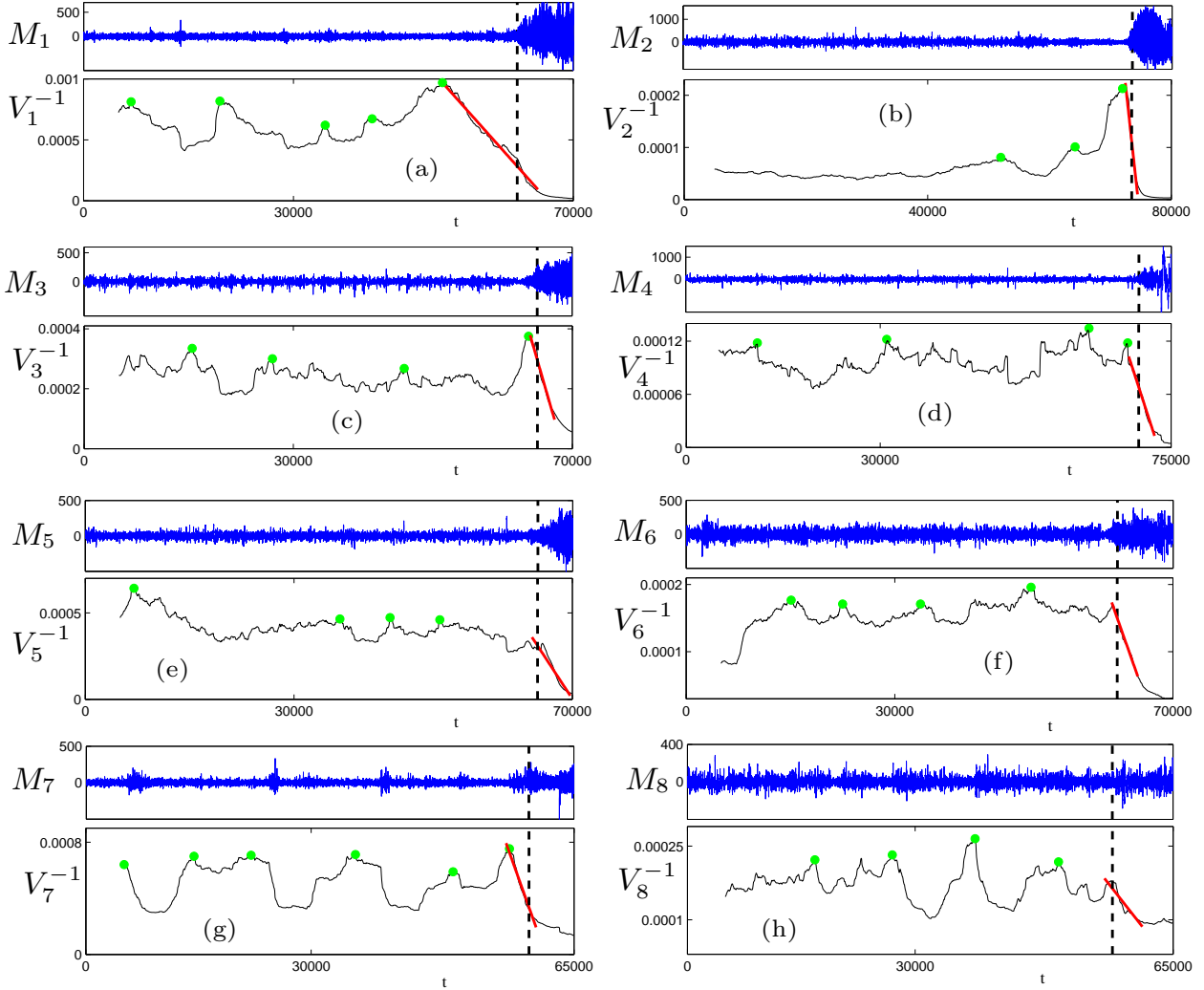}
\caption{\label{fig:fig3}The eight plots show the average channel activity $M_i$ (top, blue) and the average of the inverse variance $V_i^{-1}=1/V_i$ (bottom, black) for the eight time series $i\in\{1,2,\dots,8\}$; the horizontal axis is the time axis where the labels correspond to the sample point number. The green dots mark local maxima of $V_i^{-1}$ which correspond to local minima of $V_i$. The fitted red curves are linear and demonstrate that the variance increases near the epileptic seizure. The black dashed vertical lines are inserted for orientation purposes and roughly mark a seizure onset beyond which a very high variance is observed.}
\end{figure}

Figure \ref{fig:fig3} shows the results of this computation. We plot the inverse of the variance $V_i^{-1}$ for $i\in\{1,2,\ldots,8\}$ since this makes it easier to understand the scaling of $V_i$ near the seizure point at $t=t_c$. All four plots have some features in common:

\begin{itemize}
 \item[(A)] $V_i^{-1}$ decreases near the seizure. The scaling law seems to be given by
\be
\label{eq:Hopf_scale}
V_i\sim \frac{1}{t_c-t},\qquad \text{as $t\ra t_c$.}
\ee
 \item[(B)] There are multiple local maxima and minima for $V_i^{-1}$ before approaching the seizure point. This indicates that we should expect oscillations in statistical indicators near epileptic seizures.
 \item[(C)] The last local maximum before $t_c$ shows that there is also a period of low variance close to a seizure. 
 \item[(D)] The last local maximum before $t_c$ is already very close to the seizure. This means that predictions could be very difficult just based on a calculation of the variance.
\end{itemize}

\begin{figure}[htbp]
\centering
 \includegraphics[width=0.8\textwidth]{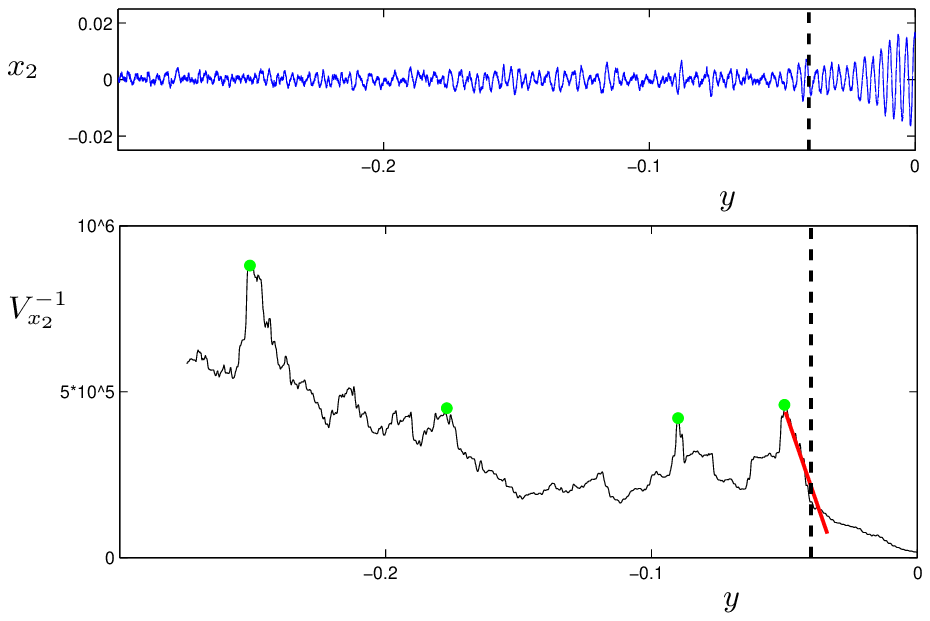}
\caption{\label{fig:fig4}Time series of the fast variable $x_2$ (top, blue) and the associated inverse of the variance $V^{-1}_{x_2}=\text{Var}(x_2(y))^{-1}$ (bottom, black) for a Hopf critical transition model \eqref{eq:Hopf} with parameter values \ref{eq:pvals-Hopf}; cf. also Figure \ref{fig:fig3}.}
\end{figure}

The next problem to consider is what types of dynamical models can reproduce the behavior we have observed from the data analysis i.e. we look for a model for the variance in clusters/regions of neurons that displays the observed oscillatory behavior and scaling law. At first glance, Figures \ref{fig:fig3}(a)-(h) could be interpreted as a summation of Figure \ref{fig:fig2}(b) i.e. of neurons that are (almost) in synchrony where the coherent spiking originates from the noise-induced escape of a spiral sink. However, the real problem in understanding the dynamical mechanism of epileptic seizures is shown in Figure \ref{fig:fig4} where we also plot the inverse of the variance $V^{-1}$ near a critical transition. The similarities to the data in Figure \ref{fig:fig3} are clear; all four observations (A)-(D) also apply in Figure \ref{fig:fig4}. The data in Figure \ref{fig:fig4} have been generated using a simple model for a Hopf critical transition \cite{KuehnCT1,KuehnCT2}:
\be
\label{eq:Hopf}
\begin{array}{rcl}
\epsilon \dot{x}_{1,\tau} &=& y_\tau x_{1,\tau}-x_{2,\tau}+x_{1,\tau}(x_{1,\tau}^2+x_{2,\tau}^2)+\sigma( a_{11}\xi_{1,\tau}+a_{12}\xi_{2,\tau}),\\
\epsilon \dot{x}_{2,\tau} &=& x_{1,\tau}+y_\tau x_{2,\tau}+x_{2,\tau}(x_{1,\tau}^2+x_{2,\tau}^2)+\sigma( a_{21}\xi_{1,\tau}+a_{22}\xi_{2,\tau}),\\
\dot{y}_\tau&=& 1,\\
\end{array}
\ee
where $\xi_{j,\tau}$ are independent white noise processes that satisfy $\langle\xi_{j,\tau_1}\xi_{j,\tau_2}\rangle=\delta(\tau_1-\tau_2)$ for $j=1,2$. The model \eqref{eq:Hopf} was first analyzed in the context of delayed Hopf bifurcation \cite{Neishtadt1,Neishtadt2}. Observe that the deterministic part of the fast variables $(x_1,x_2)$ is the normal form of a (subcritical) Hopf bifurcation \cite{Kuznetsov}. The slow variable $y$ can also be viewed as time since $y_\tau=(\tau-\tau_0)+y_0$. For the simulation in Figure \ref{fig:fig4} we have chosen
\be
\label{eq:pvals-Hopf}
\epsilon=0.0005,\qquad\sigma=0.001\sqrt\epsilon,\qquad a_{11}=1=a_{22},\qquad a_{12}=0.2=a_{21}
\ee
with a deterministic initial condition $(x_{1,0},x_{2,0},y_0)=(0,0,-0.3)$. It is known that near a sub- or supercritical Hopf bifurcation a scaling law of the form \eqref{eq:Hopf_scale} holds \cite{KuehnCT2}. Obviously the scales differ between Figure \ref{fig:fig3} and Figure \ref{fig:fig4} but those can be re-scaled to match. Therefore we have found a dynamical model that could potentially explain the qualitative features of a single variance time series for a cluster of neurons.\\

It is very important to note that we have obtained the conjecture that a Hopf bifurcation is involved in the transition to a seizure without a detailed biophysical model. In fact, several mean-field models for various types of epileptic seizures do exhibit Hopf bifurcations \cite{SuffczynskiKalitzinLopesdaSilva,Breakspearetal,Rodriguesetal,Rodriguesetal1,Martenetal} that form a boundary between a regular equilibrium (non-seizure) an oscillatory (seizure) regime. However, there are several mean-field models available \cite{Decoetal} and also other bifurcation mechanisms have been identified to play a role near seizure onset \cite{TaylorBaier}.\\

Our methods also have another important implication regarding the distinction between a preictal and a proictal state \cite{Suffczynskietal}. From a dynamical perspective, it was suggested that one can differentiate between models that show a distinct preictal state with a parameter driving the system to a bifurcation or systems showing a proictal state where noise-induced escapes play a dominant role \cite{LopesdaSilvaetal}. A subcritical Hopf bifurcation is a model that can interpolate between the two cases. Consider \eqref{eq:Hopf} in the following two cases:
\begin{enumerate}
 \item[(1)] $\epsilon=0$, $\delta>0$ and $y<0$: the equilibrium $(x_1,x_2)=(0,0)$ is a stable focus for the deterministic dynamics but it is well-known \cite{FreidlinWentzell} that a finite-time noise-induced escape always occurs. This can be viewed as the transition beyond a basin boundary given by the unstable limit cycles \cite{Suffczynskietal}. If we include another (seizure-state) attractor beyond this basin boundary we can view the situation near $(x_1,x_2)=(0,0)$ as a ``purely proictal'' state. It is well-known how to calculate the probabilistic likelihood of this Hopf transition and also for many other bifurcations involving metastability \cite{HaenggiTalknerBorkovec,Gardiner}.
 \item[(2)] $\epsilon>0$, $0<\delta\ll1$: If the noise is sufficiently small then we will reach the Hopf bifurcation point with high probability \cite{BerglundGentz} and our prediction method via scaling of the variance and critical transitions applies. We are in a ``purely preictal'' situation. 
\end{enumerate}
 
Obviously there is a continuum of possibilities in between these two situations \cite{MuratovVanden-EijndenE,BerglundGentz} depending on the scaling of noise and time scale separation. In fact, the results shown in Figure \ref{fig:fig2} illustrate the variation in such a continuum situation for the saddle-node bifurcation. A similar study of intermediate regimes for all bifurcations, including the Hopf bifurcations on a mean-field level, can certainly be carried out similar to the strategy employed in \cite{KuehnCT2}. This study is beyond the scope of this paper as we focus on the multiscale features and basic dynamical analysis techniques here.

\section{Correlations between clusters}
In the preceding two sections we investigated neuronal dynamics at different spatial scales, from single model neurons to neurophysiological data from clusters of neurons, using the variance as a univariate measure. In the following section, we will focus on the dynamics from many clusters of neurons encompassing a larger spatial scale. In contrast to the previous, bivariate measures for the activity between different clusters will be used.

Bivariate measures can take into account the correlation of two signals. Information about the correlation of neuronal activity between different anatomical regions can give insights into the state of the network as a whole. With regard to epilepsy, correlation based measures such as mean phase coherence (MPC) and maximum linear cross correlation (MLCC) have yielded promising results in identifying pre-ictal states \cite{MormannLehnertzDavidElger, Mormannetal1, Schelteretal, Feldwisch-Drentrupetal}.

In this section, we will start by considering the maximum linear cross correlation for the ECoG data used in the preceding parts, reviewing and confirming some recent observations. We will then continue to extend the bivariate analysis to wavelet-based synchronization measures able to resolve pairwise correlations at different frequency bands. We will compare these results to those obtained using MLCC and MPC. Our focus is again on multiscale character of the system with the goal of identifying scaling relationships at each level of observation.

\subsection{Maximum linear cross-correlation}

\begin{figure}[htbp]
\begin{center}
\includegraphics[width=1\textwidth]{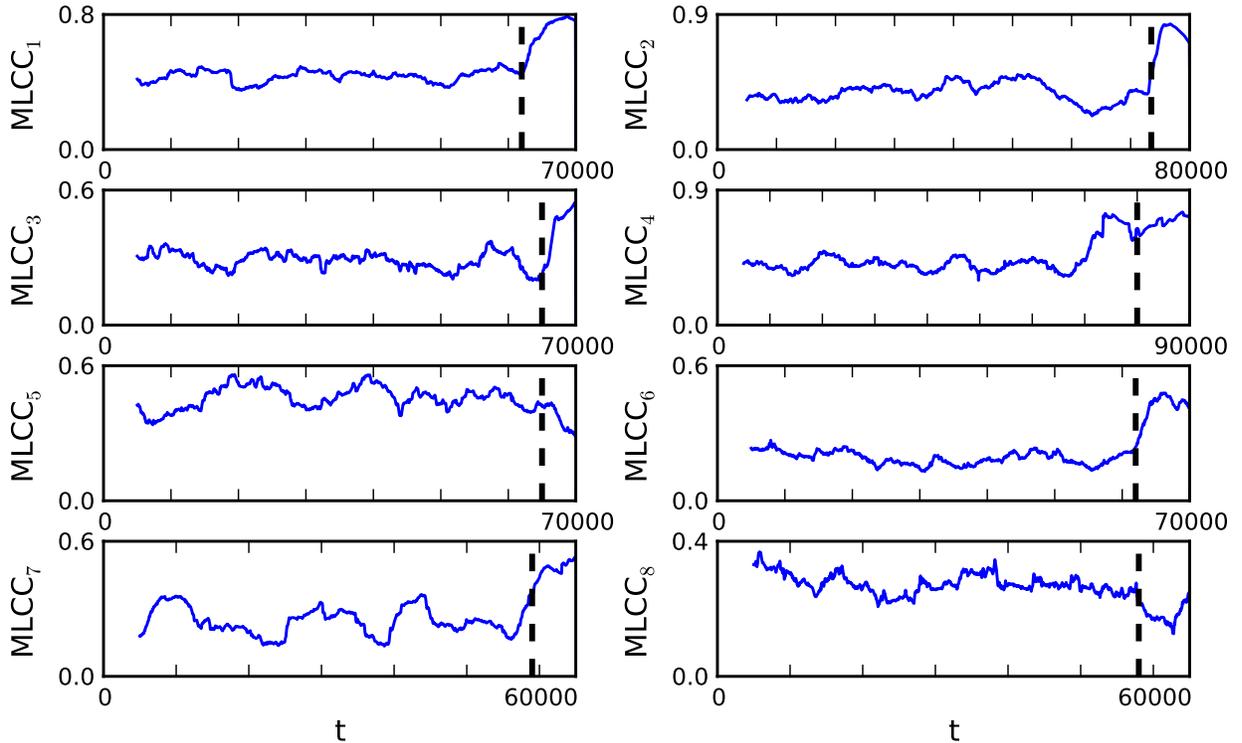}
\caption{\label{fig:fig5}Maximum linear cross-correlation $MLCC_i$ for eight pre-ictal time series $i\in\{1,2,...,8\}$. Vertical lines indicate the approximate onset of the seizure attack.}
\end{center}
\end{figure}

The maximum linear cross-correlation (MLCC) quantifies the similarity between two time series $F_i(t)$ and $F_j(t)$. MLCC is a linear measure of lag-synchronization which captures the normalized product of two time series dependent on a lag $\tau$ \cite{RosenblumPikovskyKurths}:
\begin{equation}
\label{Cmax}
C_{max}=\max_{\tau}|\frac{C_{F_iF_j}(\tau)}{\sqrt{C_{F_iF_i}(0)C_{F_jF_j}(0)}}|,
\end{equation}
where
\begin{equation}
\label{crosscorr}
C_{F_iF_j}(\tau)=\frac{1}{N-\tau}\sum^{N-\tau}_{t=1}F_i(t+\tau)F_j(t)
\end{equation}
is the linear cross-correlation function. As a measure of synchronization between activity in different anatomical areas, MLCC has been proposed and successfully applied as a precursor for pre-ictal brain activity \cite{Mormannetal1, Mormannetal2}. We computed the MLCC of 5 randomly chosen signal pairs for each time window (5000 sampling steps, a consecutive time window being shifted 50 sampling steps forward). Figure \ref{fig:fig5} shows the average over the 5 pairs for each of the 8 time series considered in the preceding sections.\\ 

\begin{figure}[htbp]
\centering
\includegraphics[width=1\textwidth]{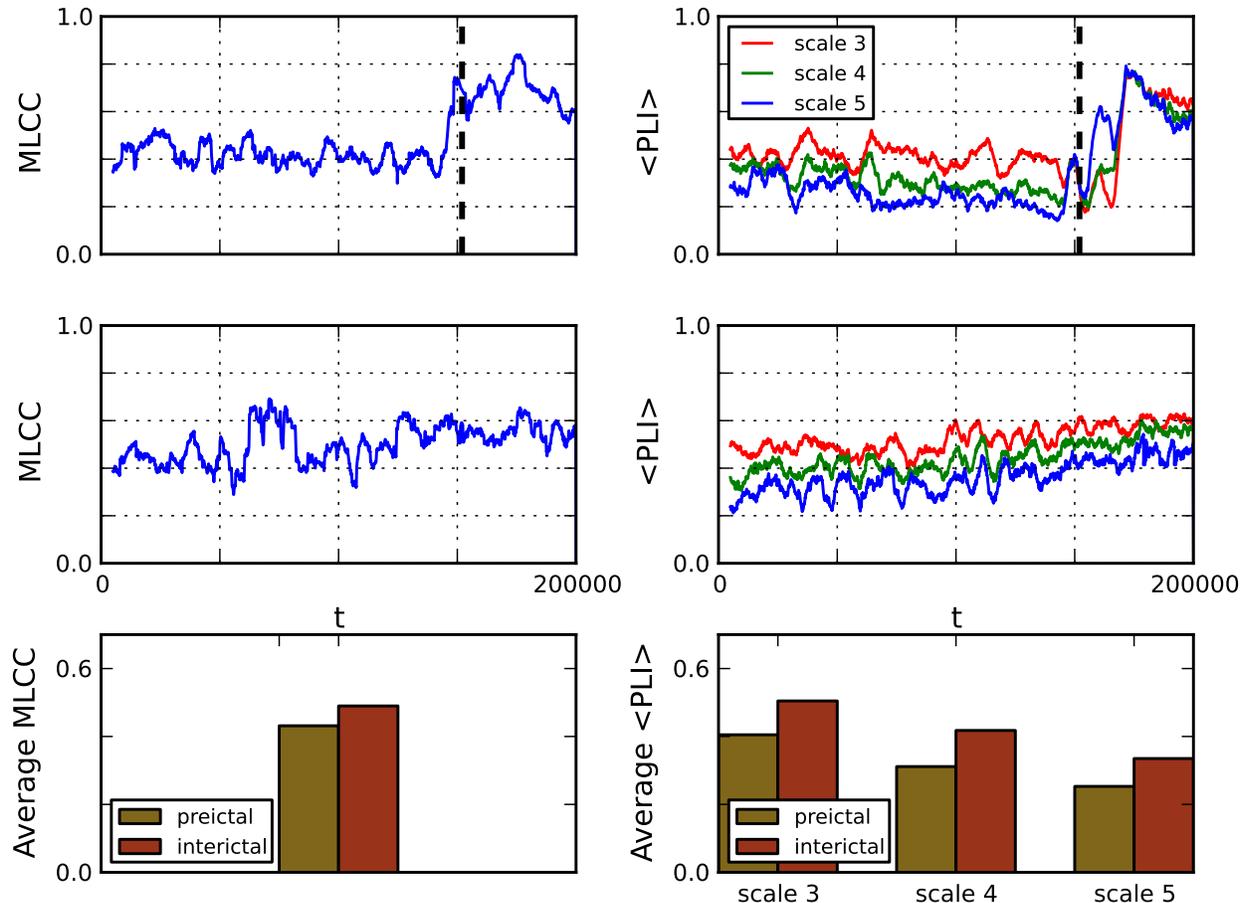}
\caption{\label{fig:fig6}Decrease of synchronization measures during a pre-ictal interval. Left column: time series of maximum linear cross correlation during a pre-ictal (top) and an inter-ictal (middle) interval. Right column: time series of $\langle{PLI}\rangle$ for three scales during a pre-ictal (top) and an inter-ictal (middle) period are depicted. Vertical dashed lines indicate the onset of the seizure attack. Averages over the first 150000 sample points of each time series indicate a distinct decrease of each synchronization measure during the pre-ictal interval (bottom row).}
\end{figure}

In most of the depicted time-series (patients 1, 2, 3, 4, 6, 7) an increase in the MLCC can be observed with the seizure onset (Fig.\ref{fig:fig5}) which is in agreement with the general observation that seizure attacks are characterized by an increased synchronization between cortical regions \cite{Lehnertzetal}. 
Prior to epileptic seizures a decrease in MLCC values has been reported and used to identify a preseizure state \cite{Mormannetal1, Mormannetal2}. 
To relate to these reports and later also compare MLCC to the wavelet-based synchronization measure $\langle{PLI}\rangle$ (see following section), we calculated MLCC for a pre-ictal and an inter-ictal time interval. Figure \ref{fig:fig6} (left column) depicts the time series of MLCC values of patient 4 during a pre-ictal (top) and an exemplary inter-ictal interval (middle), an interval being at least 6 hours apart from the next seizure attack. Average values of MLCC are plotted left in the bottom row illustrating the comparably lower values during pre-ictal intervals. MLCC levels are lower during the pre-ictal compared to the inter-ictal interval confirming recent reports of decreased synchronization as one characteristic precursor for a seizure.

A lot of effort has been put forward to utilize the observed synchronization drop in predicting seizure attacks, most of these works addressing the question whether it could be used to identify a preseizure state \cite{MormannLehnertzDavidElger, Mormannetal1, Mormannetal2, Chavezetal, Schelteretal, Feldwisch-Drentrupetal}. In this work we are not addressing this issue but focus on the dynamics and scaling relations of correlation measures near the seizure onset. For this purpose we extend the analysis to wavelets able to resolve correlations between clusters for different frequency bands.

\subsection{Wavelets}
Wavelet analysis has been applied in neuroscience research for some time \cite{Bullmoreetal, Bullmoreetal1}. 
Wavelet coefficients $W_k$ provide a frequency-dependent moving average over a time series which can be used to derive a time-resolved frequency-profile for the data given.
This capacity has also been made use of in the detection of seizures \cite{Subasi} and the investigationen of frequency profiles of epileptic seizures in humans and animals \cite{Bosnyakovaetal, LuijtelaarHramovSitnikovaKoronowskii}.
Wavelet analysis \cite{PercivalWalden} can also be used as an elegant tool to identify intervals of phase synchronization (or phase-locking) between neurophysiological time series. The phase definition can thereby be used for broad-band synchronization analysis or analysis of a specific frequency of interest. 

In this study, we investigated broad-band phase-locking between pairs of signals as introduced in \cite{WhitcherCraigmileBrown}. There, the original signal is decomposed with respect to multiple scales related to frequency bands of decreasing size. To derive a scale-dependent estimate of the phase difference between two time series, we follow the approach described in \cite{KitzbichlerSmithChristensenBullmore} using Hilbert transform derived pairs of wavelet coefficients \cite{WhitcherCraigmileBrown}. The instantaneous complex phase vector for two signals $F_i$ and $F_j$ is defined as: 

\begin{equation}
\label{instphasevector}
C_{i,j}(t)=\frac{W_k(F_i)^{\dag}W_k(F_j)}{|W_k(F_i)||W_k(F_j)|},
\end{equation}

where $W_k$ denotes the $k$-th scale of a Hilbert wavelet transform and $^\dag$ its complex conjugate. A local mean phase difference in the frequency interval defined by the $k$-th wavelet scale is then given by 

\begin{equation}
\label{argc}
\Delta\phi_{i,j}(t)=Arg(\overline{C_{i,j}}),
\end{equation}

with 

\begin{equation}
\label{caverage}
\overline{C_{i,j}}(t)=\frac{\langle W_k(F_i)^{\dag}W_k(F_j)\rangle}{\sqrt{\langle|W_k(F_i)|^2\rangle \langle|W_k(F_j)|^2\rangle}}
\end{equation}

being a less noisy estimate of $C_{i,j}$ averaged over a brief period of time $\Delta t=2^k8$ \cite{KitzbichlerSmithChristensenBullmore}. One can then identify intervals of phase-locking (PLI) as periods when $|\Delta\phi_{i,j}(t)|$ is smaller than some arbitrary threshold which we set to $\pi/4$ here. We denote phase-locking intervals between two signals $F_i$ and $F_j$ as $PLI_{i,j}$. To obtain a measure of frequency-specific phase-locking in a defined time window, we calculate the sum of $PLI_{i,j}$ for all pairs of signals and normalize this expression to confine the measure to the interval $[0, 1]$:
\begin{equation}
\label{pliaverage}
\langle{PLI}\rangle=\frac{1}{\binom{n_{signals}}{2}{n_{steps}}}\sum_{i,j}{PLI_{i,j}},
\end{equation}
where $n_{signals}$ is the number of signals and $n_{steps}$ the number of time steps in the time window under consideration.\\

\begin{figure}[htbp]
\centering
\includegraphics[width=1\textwidth]{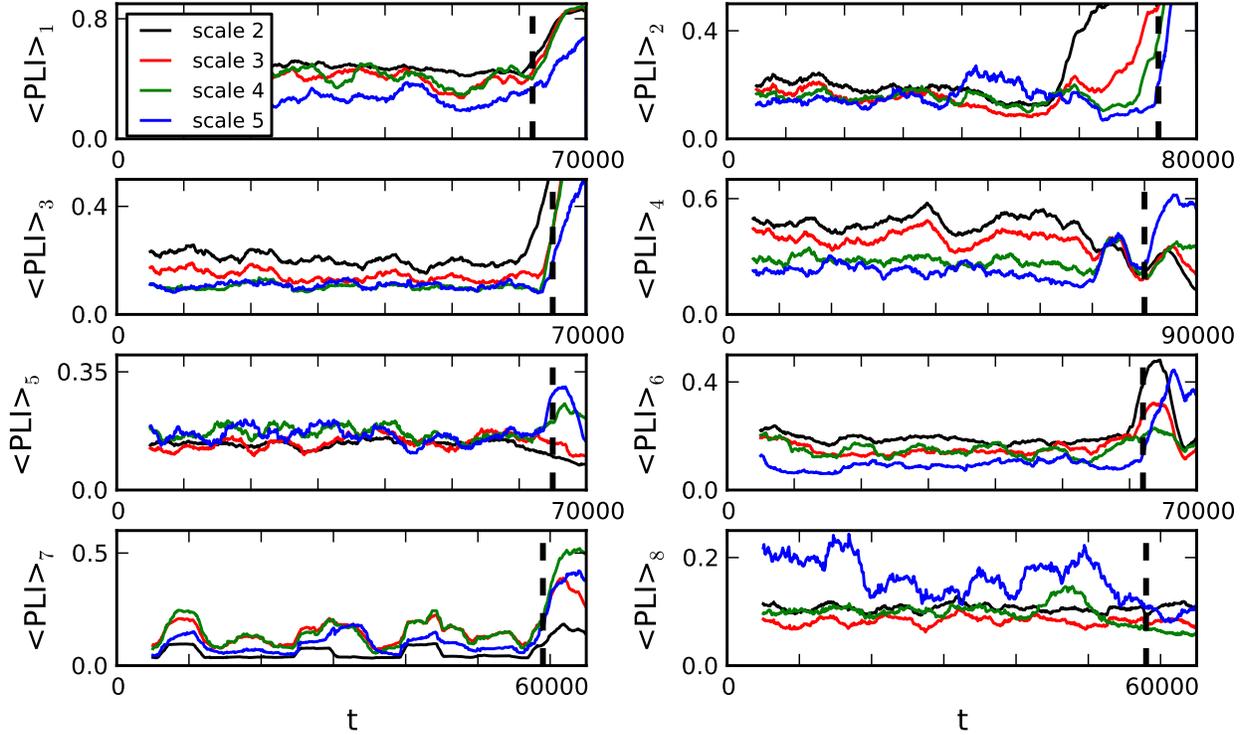}
\caption{\label{fig:fig7}Phase-locking measure $\langle{PLI}\rangle_i$ for the eight time series $i\in\{1,2,...,8\}$. Colors correspond to different scales. The vertical dashed lines indicate the approximate onset of the epileptic seizure attack.}
\end{figure}

We analyzed data for each patient for 3 different scales, referring to frequency bands 25-12, 12-6 and 6-3 Hz for patients 1-3, 5-8 and 32-16, 16-8 and 8-4 Hz for patient 4, respectively. The computation of $\langle{PLI}\rangle$ was done for time windows of 5000 sampling steps, consecutive time windows were shifted forward by 50 sampling steps. Figure \ref{fig:fig7} shows the results of this computation. 

In all 8 patients, comparably low values of $\langle{PLI}\rangle$ ($\langle{PLI}\rangle\leq0.5$) are observed for all scales. Similarly to MLCC synchronization as measured by the phase-locking intervals for different scales is decreased during an pre-ictal interval compared to an inter-ictal one (Fig. \ref{fig:fig6}, right column). The observed decrease in synchronization measure suggests that application of $\langle{PLI}\rangle$ could also prove useful in preseizure state detection algorithms, similar to the MLCC.

As mentioned earlier, our focus is on the dynamical behavior near the seizure onset. 
Aside from the aforementioned low values of $\langle{PLI}\rangle$, some characteristic features can be observed:

\begin{itemize}
\item[(A)] Phase-locking measured by $\langle{PLI}\rangle$ increases around seizure onset times. (Similar to MLCC, this is seen less clearly in patients 5 and 8.)
\item[(B)] The increase of $\langle{PLI}\rangle$ for different scales often starts at different times. Phase-locking of lower scales tends to increase earlier compared to higher scales.
\item[(C)] The increase of $\langle{PLI}\rangle$ appears to be linear.
\end{itemize}

\begin{figure}[htbp]
\centering
\includegraphics[width=1\textwidth]{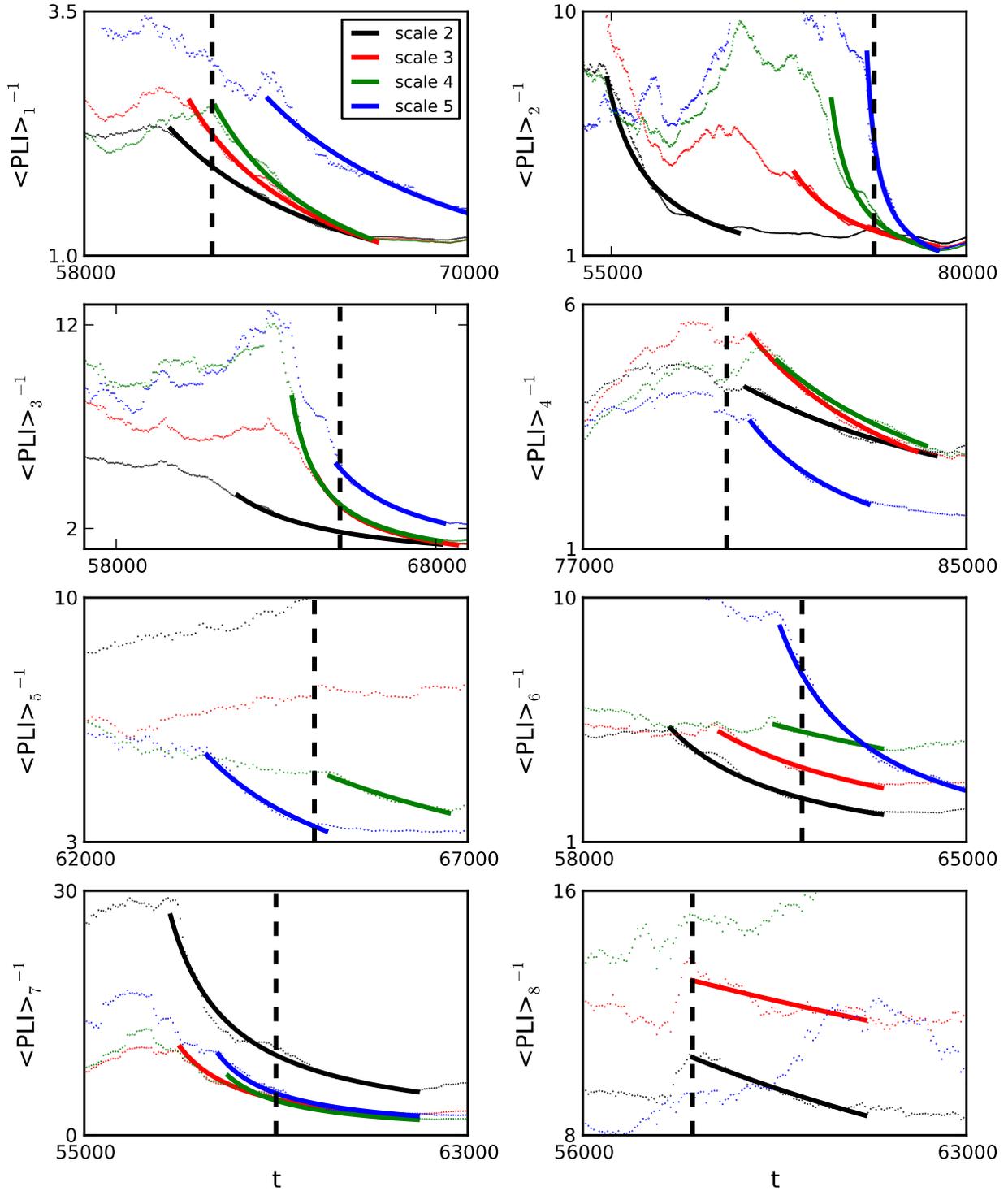}
\caption{\label{fig:fig8}The inverse $\langle{PLI}\rangle^{-1} = 1/\langle{PLI}\rangle$ for the eight time series is plotted near the seizure onset. The fitted go with $1/t$ and demonstrate that the variance increases linearly near the epileptic seizure (black dashed vertical lines).}
\end{figure}

Point A comes not as a surprise reflecting the fact of increased synchronization between cortical regions observed during seizures. For better clarity regarding point B and C, we plot the inverse of $\langle{PLI}\rangle$, $\langle{PLI}\rangle^{-1}$, for times close to the seizure onset (Fig. \ref{fig:fig8}). In many of the patients (1, 2, 3, 6, 7) one can observe that $\langle{PLI}\rangle^{-1}$ of lower scales decreases earlier than higher scales meaning that $\langle{PLI}\rangle$ of higher frequencies shows an earlier increase compared to lower frequencies. In fact, the temporal order in which $\langle{PLI}\rangle$ increases often appears to be directly correlated to the numerical order of scales (see patient 1 in fig. \ref{fig:fig8} for example). We furthermore observe $\langle{PLI}\rangle^{-1}$ to decrease with $\frac{1}{t}$ suggesting a linear increase of its inverse $\langle{PLI}\rangle$ with time. Fitting a linear function $\langle{PLI}\rangle \propto t$ close to seizure onset times also provided the better fit compared to power-law or exponential relationships (Fig. \ref{fig:fig8}).

\begin{figure}[htbp]
\centering
\includegraphics[width=1\textwidth]{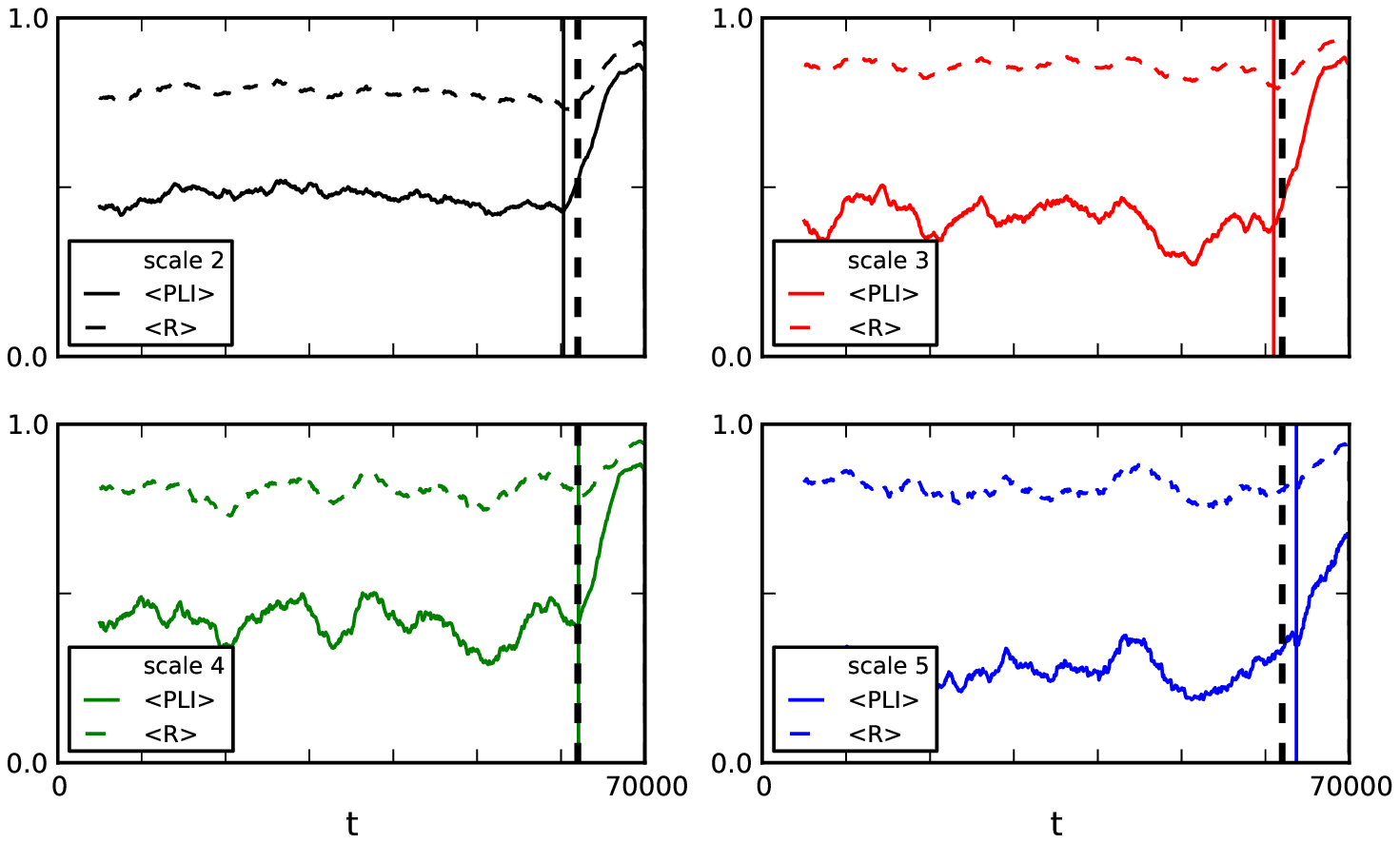}
\caption{\label{fig:fig9}Comparison between $\langle{PLI}\rangle$ and mean phase coherence $\langle R \rangle$ for patient 1. Both measures based on phase-synchronization show a similar behavior. Colored vertical lines indicate the beginning of the increase in synchronization near the seizure onset (black dashed vertical line). Both measures show an earlier increase for lower scales.}
\end{figure}

Another nonlinear measure based on phase synchronization is the mean phase coherence \cite{MormannLehnertzDavidElger, Chavezetal}. For two pairs of neurophysiological time series $F_i$ and $F_j$ it is given by 

\begin{equation}
\label{mpc}
R_{i,j}=\sqrt{\langle \cos \Delta\phi_{i,j}(t) \rangle^2 + \langle \sin \Delta\phi_{i,j}(t) \rangle^2}
\end{equation} 

with $\Delta\phi_{i,j}(t)$ being the phase difference between the two signals at time $t$ and $\langle \rangle$ denoting the average over time. We calculated $R_{i,j}$ for all pairs of signals using the wavelet-derived, scale-dependent phase differences for each patient. The average $\langle R \rangle$ over all $R_{i,j}$ showed a similar time course as $\langle{PLI}\rangle$ (Fig. \ref{fig:fig9}). Near seizure onset, the same temporal order of the increase in synchronization was observed indicating independence from the specific measure of phase-synchronization.
Direct comparison of both nonlinear synchronization measures $\langle R \rangle$ and $\langle{PLI}\rangle$ to MLCC suggests that the frequency resolved measures add new information at the onset of the seizure. Therefore such multiscale measures may potentially be better suited to explain the dynamical process that causes a seizure attack.

\section{Discussion}

In the present paper we aimed for a better understanding of the dynamical processes involved in seizure generation. Our approach extended over three spatial scales involving two recently developed methods (critical transitions and wavelet derived phase-lock intervals).
We showed for the first time that the theory of critical transitions \cite{KuehnCT1,KuehnCT2} can be applied in the context of excitable neurons operating near the spiking threshold. On the level of clusters of neurons we identified a potential Hopf bifurcation as the seizure onset mechanism from data based on this theory and found a new scaling law of single-event time series data. On the largest spatial scale we observed a scaling law occurring at time-shifted onset times and compared our wavelet-based phase-locking measure to other bivariate measures.\\

One of our main results is the observation of scaling laws on different spatial scales -- for individual neurons \eqref{eq:Var}, for activity of clusters of neurons \eqref{eq:Hopf_scale} and for the increase of phase-locking near the seizure onset. A recent publication highlighted five power-law scaling laws related to epileptic seizures and their analogy to earthquakes (the Gutenberg-Richter distribution of event sizes, the distribution of interevent intervals, the Omori and inverse Omori laws and the conditional waiting time until next event) \cite{Osorioetal1}. Other works investigating scaling laws of ictal and interictal epochs reported similar inter-seizure-interval statistics in genetically altered rats while in human data no power-law distribution was observed \cite{SuffczynskiKalitzinLopesdaSilva, Suffczynski2006}.\\

The observation of such scaling laws is important because it may guide new models of seizure dynamics by allowing insights into the dynamical processes that may have generated the underlying data.
Many of the scaling laws reported here and elsewhere \cite{Osorioetal1} exhibit power laws. The observation of similar scaling laws on different spatial scales, from single neurons to the size distribution of different seizures, strongly emphasizes the multi-level character of epileptic seizure generation. More importantly, it yields insights into the dynamical properties of the underlying system \cite{Jost}. The strong analogies between seismic shocks and brain seizures have previously been pointed out and hypothesized to emerge from the structural commonality of the two systems: both are composed of interacting nonlinear threshold oscillators and are far from equilibrium \cite{Kapirisetal}. Critical dynamics is believed to be a consequence of these structural properties in both these systems. Recent findings in preparations of rat cortex \cite{BeggsPlenz} and primate brain \textit{in vivo} \cite{Petermannetal} exhibiting power-law statistics of activity, a hallmark of phase transitions \cite{BakPaczuski, LevinaHerrmannGeisel, MeiselGross}, have led to the hypothesis that also human brain dynamics is poised at a phase transition \cite{Beggs2004, KitzbichlerSmithChristensenBullmore}. Although such statistics can result from different processes, the self-similar behavior captured by the diverse scaling laws on different levels might potentially be related to the notion of criticality in brain dynamics. Models describing epilepsy should also resemble these multi-level scaling laws and take into account critical brain dynamics.\\ 

Decomposition into different spatial scales showed oscillations in a pre-seizure state at all levels. Observation of such oscillations in real world data offers characteristics to be useful when testing future models. 
As we showed here, based on critical transition theory, the variance's oscillations along with its scaling law are generically characteristics for Hopf bifurcation. These results therefore validate previous seizure models assuming a Hopf bifurcation as a main mechanism as the transition point \cite{Rodriguesetal, SuffczynskiKalitzinLopesdaSilva, Martenetal}.
While the goal in seizure prediction is to predict large events, there is growing consensus about the key role played by small events, from precursor oscillations to subclinical seizures \cite{MormannAndzejakElgerLehnertz, Osorioetal1}. 
Future models and predictor systems should encompass those as prediction algorithms unable to account for such small oscillations would be ill-adapted and likely provide incorrect seizure forecasts.\\

Near seizure onset we observed a time shifted increase in phase-locking; phase-locking for higher frequencies (lower scales) tended to precede lower frequencies (higher scales). In a recent study, wavelet analysis of spike-wave discharges, a different form seizure activiy, revealed changes in the time-frequency dynamics during discharges. While initially a short period with the highest frequency value was observed, the frequency later decreased \cite{Bosnyakova2006, Bosnyakovaetal}. Other studies showed high frequency oscillations specifically at seizure onset \cite{Wendlingetal, Molaee-ArdekaniBenquetBartolomeiWendling}, see \cite{Richardson} for a comprehensive overview. Together these studies demonstrate dynamic changes in the time-frequency domain of seizures with higher dominating frequencies at seizure onset. Our observations complement these findings suggesting a frequency-dependent, shifted start of synchronization near seizure onset.\\

\appendix

\section{Data}
\label{ap:data}

Eight patients undergoing surgical treatment for intractable epilepsy participated in the study. Patients underwent a craniotomy for subdural placement of electrode grids and strips followed by continuous video and electrocorticogram (ECoG) monitoring to localize epileptogenic zones. Solely clinical considerations determined the placement of electrodes and the duration of monitoring. All patients provided informed consent. The study protocols were approved by the Ethics Committee of the Technical University Dresden. ECoG signals were recorded by the clinical EEG system (epas 128, Natus Medical Incorporated) and bandpass filtered between $0.53$ Hz and $70$ Hz. Data were continuously sampled at a frequency of $200$ Hz (patients $1-3$ and $5-8$) and $256$ Hz (patient $4$, \cite{Ihleetal}). We always indicate the sampling point number on the time axis if we use the data. No claims regarding a large-scale statistical validity of the data set is made since the total patient sample size is rather small. Although this is an important issue \cite{Schelteretal} we focus here on identifying the dynamical mechanisms and new time series analysis techniques in the context of epileptic seizures.

\end{document}